\documentclass[prl,superscriptaddress,twocolumn]{revtex4}
\usepackage{epsfig}
\usepackage{latexsym}
\usepackage{amsmath}
\begin {document}
\title {Periodicity of mass extinctions without an extraterrestrial cause}
\author{Adam Lipowski}
\affiliation{Faculty of Physics, Adam Mickiewicz University,
61-614 Pozna\'{n}, Poland}
\begin {abstract}
We study a lattice model of a multi-species prey-predator system.
Numerical results show that for a small mutation rate the model
develops irregular long-period oscillatory behavior with sizeable
changes in a number of species. The periodicity of extinctions on
Earth was suggested by Raup and Sepkoski but so far is lacking a
satisfactory explanation. Our model indicates that this might be a
natural consequence of the ecosystem dynamics, not the result of
any extraterrestrial cause.
\end{abstract}
\maketitle
The Earth ecosystem is certainly a subject of intensive
multidisciplinary research. Researchers in this field believe that
at least some basic understanding of this immensely complex system
can be obtained using relatively simple models, that nevertheless
grasp some aspects of its rich behavior~\cite{LV}. Of particular
interest in physicists community is the dynamics of extinctions of
species~\cite{NEWMAN}. Palaeontological data, that show broad
distributions of these events in the Earth history, suggest
existence of strong, perhaps power-law correlations between
extinctions. Similar correlations appear in the so-called critical
systems and such an analogy resulted in a wealth of interesting
models that consider extinctions as a natural consequence of the
dynamics of an ecosystem~\cite{BAKSNEPP}. However, fossil data are
not entirely convincing, and it is not clear to what extent the
analogy with critical systems hold. What is more, a number of
researchers prefer an alternative explanation where extinctions
appear due to external stresses imposed on the ecosystem as, e.g.,
impacts of comets or meteorites, or an increased volcanic
activity~\cite{NEWMAN}. The popularity of theories of exogenous
origin of extinctions increased when Raup and Sepkoski concluded
from analyzing fossil date that big extinction events during the
last 250 My (million years) have been occurring with periodicity
of about 26 My~\cite{raupsep}. Several theories, mostly of
astronomical origin, have been proposed to explain such a
periodicity, but none of them is confirmed or commonly
accepted~\cite{theories}. Although the Raup and Sepkoski analysis
was put into question~\cite{patterson}, the more recent analysis
confirms a similar periodicity of extinctions~\cite{prokoph}
keeping this fascinating hypothesis still open.

Lacking a firm evidence of any exogenous cause, one can ask
whether the periodicity of extinctions can be explained without
referring to such a factor. Or in other words, if it is possible
that the ecosystem dynamics produces (by itself) oscillations on
such a long time scale. Since the seminal work of Lotka and
Volterra, an oscillatory behavior is already well-known in various
prey-predator systems~\cite{LV,SZABO}, but the periodicity of
oscillations of densities in such systems, that is determined by
the growth and death rate coefficients of interacting species, is
of the order of a few years rather than millions. Prey-predator
systems, where such an oscillatory behavior was studied, are
typically quite simple and consist of a fixed and rather small
number of species. Certainly a model capable of describing the
dynamics of extinctions should include a large number of species
as well as mutation and competition mechanisms. There is already a
wealth of papers where various models of this kind where
examined~\cite{ZIA}, but none of them has been reporting a
long-term periodicity of extinctions. There is, however, one
aspect that these models are missing and that is perhaps quite
important, namely they neglect spatial correlations between
organisms. From statistical mechanics we already know that when
the spatial dimension of the embedding space is rather low, such
correlations might play an important role, and hence more
realistic models of the ecosystem should take them into account.

In the present paper we study  a multi-species lattice model of an
ecosystem. In our model predator species compete for food (prey)
and space (to place an offspring). This competition combined with
a mutation mechanism leads to the periodic behavior, although some
characteristics of our model, as, e.g., the number of species,
show in addition strong stochastic irregularities. Sometimes our
system is populated by a group of medium-efficiency species. But
this coexistence at a certain moment is interrupted by creation of
a species that is more efficient and able to invade even a
substantial part of the system. However, the reign of such an apex
predator does not last long. It is a fast-consuming species and it
quickly decimates the population of preys, which in turn leads to
its own decline. Such a situation opens up niches that again
become occupied by less-effective species that survived the
invasion or were created by mutation, and the situation repeats.
Simulations show that the smaller the mutation probability, the
larger the periodicity of such a behavior. Although it is
difficult to access, we expect that the mutation rate in real
ecosystems, as interpreted in the context of our
model~\cite{comm1}, is very small and the presented model might at
least suggest an explanation of the 26My periodicity of big
extinctions as a natural consequence of the ecosystem dynamics,
not as the result of an external perturbation.

Our model is a multi-species extension of an already examined
prey-predator model~\cite{lip1999}. At each site $i$ of a square
lattice of linear size $N$ we have the four-state operator $x_i$
that corresponds to this site being empty ($x_i=0$), occupied by a
prey ($x_i=1$), by a predator ($x_i=2$), or by both of them
($x_i=3$). Each predator is characterized by a real number
parameter $m_i$ ($0<m_i<1$) that we will call size ($m_i$ is
meaningful only when $i$ is occupied by a predator). We also
introduce the relative update rate of preys and predators $r$
($0<r<1$), and the mutation
probability $p$. The dynamics of this model is specified as follows:\\
(a) Choose a site at random (the chosen site will be denoted by $i$).\\
(b) With the probability $r$ update a prey at site $i$ (i.e., if
$x_1=1$ or $x_i=3$, otherwise do nothing). Provided that at least
one neighbor (say $j$) of the chosen site is not occupied by a
prey (i.e., $x_j=0$ or $x_j=2$), the prey at the site $i$ produces
an offspring and places it on an empty neighboring site (if there
are more empty sites, one of them is chosen randomly). Otherwise
(i.e., if there are no empty sites) the prey does not breed.\\
(c) Provided that $i$ is occupied by a predator (i.e., $x_1=2$ or
$x_i=3$) update this site with the probability $(1-r)m_i$, where
$m_i$ is the size of the predator at site $i$. If a chosen site is
occupied by a predator only ($x_i=2$), it dies, i.e., the site
becomes empty ($x_i=0$). If there is also a prey there ($x_i=3$)
the predator consumes the prey (i.e., $x_i$ is set to 2) and if
possible it places an offspring at an empty neighboring site. For
a predator of size $m_i$ it is possible to place an offspring at a
site $j$ provided that $j$ is not occupied by a predator ($x_j=0$
or $x_j=1$) or is occupied by a predator ($x_j=2$ or $x_j=3$) but
of a smaller size than $m_i$ (in such a case a smaller-size
predator is replaced by an offspring of a larger-size predator).
An offspring inherits parent's size with the probability $1-p$ and
with the probability $p$ it gets a new size that is drawn from a uniform distribution.\\

One can see that the size $m_i$ of a predator determines both its
update rate and its strength when it competes with other predators
for space. While the increased strength is always favorable, the
larger update rate might be a disadvantage when preys do not
reproduce fast enough. As it will be shown below, the behavior of
our model is very much influenced  by this property of the
dynamics.

The already studied single-predator version~\cite{lip1999} is
obtained when all predators have a unit size $m_i=1$ and
suppressed mutations $p=0$. In such a case, for $r>0.11$ the model
is in an active phase with positive densities of preys $\rho_0$
(which is a fraction of all sites $i$ such that $x_i=1$ or
$x_i=3$) and predators $\rho$ (fraction of all sites $i$ such that
$x_i=2$ or $x_i=3$). For $r<0.11$ the update rate of preys is too
small to sustain an active phase but it is a population of
predators that becomes extinct and the model enters an absorbing
state where all sites are occupied by preys. In the active phase
but close to the transition point (0.11) one observes oscillations
of $\rho_0$ and $\rho$ but the amplitude of these oscillations
diminishes in the thermodynamic limit $N\rightarrow\infty$. On the
other hand, for the model on the three-dimensional lattice such
oscillations most likely persist in this limit~\cite{lip1999}.

\begin{figure}
\vspace{-2cm} \centerline{ \epsfxsize=8cm \epsfbox{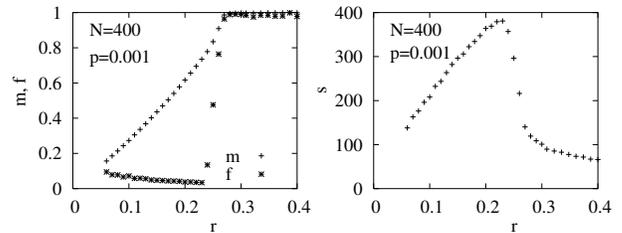} }
\caption{Average size $m$, fraction of a dominant predator species
$f$ and the number of species $s$ as a function of update rate
$r$. Results of simulations do not depend on an initial
configuration and usually it was a random distribution of preys
and predators.} \label{steady}
\end{figure}

To examine the behavior of our model we used simulations and
measured its various characteristics such as densities of preys
$\rho_0$ and predators $\rho$, the average size of dominant
predator $f$, the average size $m$, the number of species $s$, and
the lifetime of predator species. To define $s$ we classify
predators into species according to their size. Some of these
quantities are presented in Fig.~\ref{steady}. One can see that
for $r>r_c\sim 0.27$ predators in the system belong essentially to
one dominant ($f\sim 1$) species of a large size ($m\sim 1$). Of
course, mutations create from time to time some other species but
they occupy a negligible portion of a system -- unless a newly
created species will have a larger size than the dominant species
and will be able to invade the system. Fig.~\ref{steady} also
shows that a much different behavior appears  for $r<r_c$. In this
case a dominant species occupies only a small fraction of a system
(the comparison with the results for system size $N=200$ shows a
strong $N$-dependence and suggests that for larger $N$ the
fraction $f$ will diminish to zero). Moreover, the average size
$m$ differs substantially from unity that indicates that having a
large size is no longer advantageous feature. Another indication
of a more complex behavior in this case is a large increase of the
number of species.

\begin{figure}
\centerline{ \epsfxsize=9cm \epsfbox{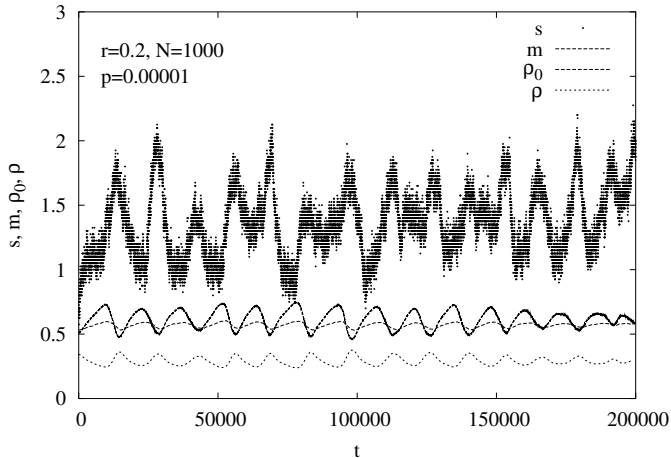} }
\caption{Time dependence of the number of species $s$ (to
superpose with other data it was divided by 40), average size $m$,
density of preys $\rho_0$, and density of predators $\rho$.}
\label{time00001}
\end{figure}

In our opinion, it is the regime for $r<r_c$ whose complex
dynamics might resemble the behavior of realistic ecosystems. To
have a better understanding of the behavior of the model in this
regime we present a time dependence of some of its
characteristics. The unit of time is defined as a single, on
average, update of each site (i.e., it is made of $N^2$ elementary
single-site updates). While in Fig.~\ref{time00001} densities
$\rho_0$, $\rho$, and the average size $m$ show a relatively
regular oscillations, the number of species $s$ is much more
irregular. During periods of multi-species coexistence, predators
have a rather small size (they eat slowly) that enables them to
sustain their density $\rho$ relatively large. As a result density
of preys $\rho_0$ is rather small. At certain moment, however, a
predator of a large size is created and starts to invade the
system. As a result the number of species $s$ rapidly decreases
while $m$ increases. Moreover, the density $\rho$ decreases and
this is related with the fact that a predator of a large size
consumes preys too quickly and is simply running out of food.
Hence, the population of this predator in some places disappears
and that creates areas where preys can breed without being
consumed by predators and that is why the density $\rho_0$ after
an initial short decline increases to a relatively large value.
However, a large-size species cannot keep its dominance for a long
time since large empty places occupied mainly by preys constitute
ideal niches for other predators as well. As a result, the model
is driven again toward a multispecies coexistence.

\begin{figure}
\centerline{ \epsfxsize=9cm \epsfbox{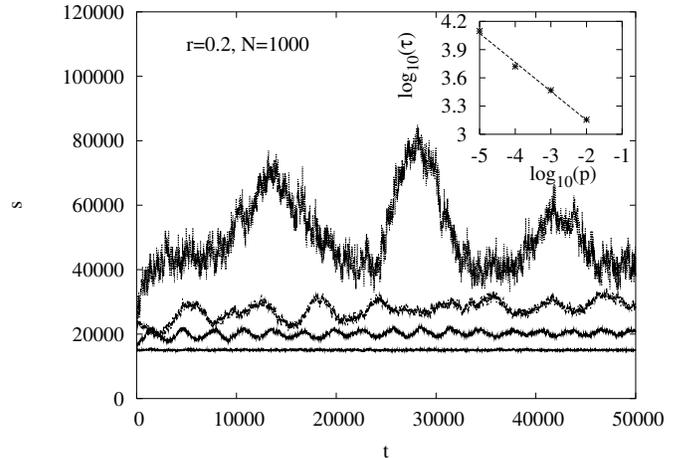} }
\caption{The time dependence of the number of species for (from
top) $p=0.00001$, 0.0001, 0.001, and 0.01. To superpose the data
on a graph the actual values of $s$ were divided by some factors.
Such an operation does not change a characteristic period of
fluctuations and their relative amplitude. Inset shows the period
of oscillation $\tau$ as a function of mutation probability $p$
obtained from the maximum of the Fourier transform of the time
dependence of the number of species ($N=1000$)}
\label{mutr02n1000}
\end{figure}

An important question is how these oscillations behave for an
increasing system size $N$. Comparing (not presented) results for
different values of $N$, we expect that the amplitude of these
oscillations will diminish to zero (period of oscillations does
not seem to depend on $N$). This is because for a sufficiently
large $N$ the system is essentially decomposed into several
independent domains where multi-species and fewer-species periods
are uncorrelated and fluctuations cancel out. However, there is an
additional factor that is responsible for the size of these
independent domains and thus the amplitude of oscillations, namely
the mutation probability $p$. Indeed, the end of the multi-species
period in a certain domain is induced by the creation of a
large-size predator. For the decreasing mutation probability $p$
such events will be less and less frequent and multi-species
domains will have more time to grow. We thus expect that for
decreasing $p$ the size of such domains should increase and, as a
result, for finite $N$ the amplitude of oscillations should also
increase. Moreover, the period of these oscillations, that is
determined by the time needed for such domains to grow, should
also increase. Simulations, as shown in Fig.~\ref{mutr02n1000},
confirm such a behavior. Let us notice large fluctuations for
$p=0.00001$, where the number of species after an invasion drops
roughly by a factor of two. To examine the $p$-dependence of the
period of oscillations $\tau$ more quantitatively, we calculated
the Fourier transform of the time dependent number of species $s$
(other characteristics like $m$, $\rho_0$, or $\rho$ give
basically the same result). The period of oscillations $\tau$ as
extracted from the maximum of this transform is shown on the
logarithmic scale in the inset of Fig.~\ref{mutr02n1000}. Straight
line fit corresponds to the dependence $\tau\sim p^{-0.31}$ but
calculations for larger system size $N$ or smaller $p$ might
modify this estimation.

As we already mentioned, the amplitude of oscillations in our
model is determined by the combination of two factors: system size
$N$ and mutation probability $p$.  That $\tau$ increases for
decreasing $p$ is an important result. It shows that for a small
mutation probability $p$ and finite but large system size $N$
(i.e., specifications of the real ecosystem) the model develops
long-period oscillatory behavior with sizeable changes of e.g.,
the number of species $s$. It might be interesting to notice that
for the single-predator version~\cite{lip1999}, with $m_i=1$ and
$p=0$, the period of oscillations in the two-dimensional case for
$r=0.2$ is around $30$. For the present model and for $p=0.00001$
the period of oscillations is larger by almost three orders of
magnitude (see the inset in Fig.~\ref{mutr02n1000}). It shows that
the periodic behavior in our model has a much different mechanism
than the Lotka-Volterra oscillations in simple prey-predator
systems.

\begin{figure}
\centerline{ \epsfxsize=9cm \epsfbox{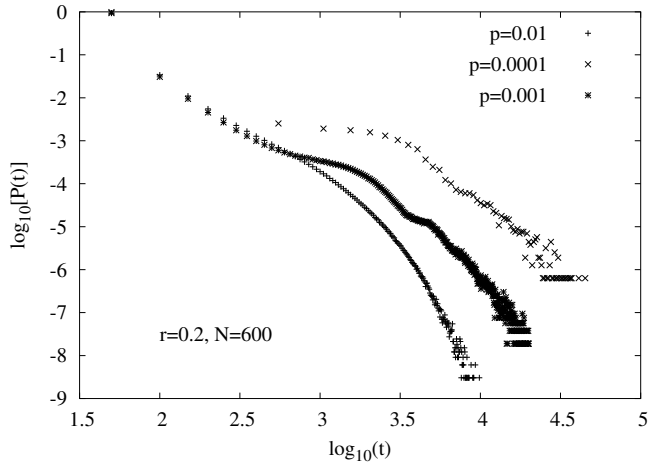} }
\caption{Logarithmic plot of the probability distribution of
lifetimes of predator species.} \label{tau}
\end{figure}

One of the properties often analyzed in models of ecosystems  is
the lifetime distribution of species. Palaeontological data
suggest some broad distributions but they are again not very
conclusive and both exponential and power-law fits can be
made~\cite{NEWMAN}. The lifetime distribution for our model is
shown in Fig.~\ref{tau}. Although for $p=0.01$ the numerical
results suggest an exponential distribution, for smaller $p$ the
situation is less clear. Especially, for $p=0.00001$ it seems that
a broader, perhaps a power-law distribution might better describe
the lifetime of our species.

It would be interesting to make further analysis of our model. For
example, one can implement a less abrupt mutation mechanism, where
a new species will be only a small mutation of a parental species.
Such a modification probably results in a longer period of
oscillations and might be more suitable for comparison with the
real ecosystem. Another possibility might be to examine the
differences in, e.g., lifetime distribution of species before and
after an extinction. That such differences exist is suggested by
the asymmetry of our data in Fig.~\ref{time00001}, where the
changes in a pre-extinction period seem to be different than in
the post-extinction one. Palaeontological data also show certain
differences in longevity of species during such
periods~\cite{Miller}, and a comparison with the predictions of
our model, if feasible, would be very desirable.

Acknowledgements: I thank M.~Droz (Univ. of Geneva) and
A.~P\c{e}kalski (Univ. of Wroc\l aw) for interesting discussions.
The research grant 1 P03B 014 27 from KBN is gratefully
acknowledged.


\begin{thebibliography}{}
\bibitem{LV} J.~D.~Murray, {\it Mathematical Biology}, (Springer, 1989).
J.~Hofbauer and K.~Sigmund {\it The Theory of Evolution and
Dynamical Systems}, (Cambridge University Press, 1988).
\bibitem{NEWMAN} M.~E.~J.~Newman and R.~G.~Palmer, {\it Modelling Extinction}, (Oxford University Press, New York, 2003).
M.~E.~J.~Newman and R.~G.~O.~Palmer, e-print: adap-org/9908002.
\bibitem{BAKSNEPP} P.~Bak and K.~Sneppen, Phys.~Rev.~Lett.~{\bf 71}, 4083 (1993).
R.~V.~Sol\'e and S.~C.~Manrubia, Phys.~Rev.~E {\bf 54}, R42
(1996). L.~A.~N.~Amaral and M.~Meyer, Phys.~Rev.~Lett.~{\bf 82},
652 (1999). B.~Drossel, Adv.~Phys.~{\bf 50}, 209 (2001).
\bibitem{raupsep} D.~M.~Raup and J.~J.~Sepkoski, ~
Proc.~Natl.~Acad.~Sci.~{\bf 81}, 801 (1984).
\bibitem{theories} M.~Davis, P.~Hut, and R.A.~Muller, Nature {\bf
308}, 715 (1984). M.~R.~Rampino and R.~B.~Stothers, Nature {\bf
308}, 709 (1984). D.~P.~Whitmire and A.~A.~Jackson, Nature {\bf
308}, 713 (1984).
\bibitem{patterson} C.~Patterson and A.~B.~Smith, Nature {\bf
330}, 248 (1987). S.~M.~Stanley, Paleobiology {\bf 16}, 401
(1990).
\bibitem{prokoph} A.~Prokoph, A.~D.~Fowler, and R.~T.~Patterson,
Geology {\bf 28}, 867 (2000).
\bibitem{SZABO} Various kinds of oscillatory behaviour without a
periodic driving are known in for example epidemic spreading
(M.~Kuperman and G.~Abramson, Phys.~Rev.~Lett.~{\bf 86}, 2909
(2001)) or voter-like models (A.~Szolnoki and G.~Szab\'o,
Phys.~Rev.~E {\bf 70}, 037102 (2004)).
\bibitem{ZIA} P.~A.~Rikvold and R.~K.~P.~Zia, Phys.~Rev.~E {\bf 68}, 031913 (2003). D.~Chowdhury,
D.~Stauffer, and A.~Kunwar, Phys.~Rev.~Lett.~{\bf 90}, 068101
(2003). B.~Drossel and A.~J.~McKane, {\it Handbook of Graphs and
Networks: From the Genome to the Internet}, S.~Bornholdt and
H.G.~Schuster (Eds) (Wiley-VCM, Berlin, 2002), e-print:
nlin.AO/0202034. M.~Hall, K.~Christensen, S.~A.~di Collobiano, and
H.~J.~Jensen, Phys.~Rev.~E {\bf 66}, 011904 (2002). C.~Quince,
P.~G.~Higgs, and A.~J.~McKane, in {\it Biological Evolution and
Statistical Physics}, eds. M.~L\"{a}ssig and A.~Vallerian
(Springer Verlag, Berlin-Heidelberg 2002). F.~Coppex, M.~Droz, and
A.~Lipowski, Phys.~Rev.~E \textbf{ 69}, 061901 (2004).
\bibitem{comm1} Although mutations are relatively frequent, for
comparison with our model we would have to consider events that
from a given species create a species that is 'substantially'
different. Such events usually result as a cumulative effect of
many mutations and their probability is certainly very small.
\bibitem{lip1999} A.~Lipowski, Phys.~Rev.~E {\bf 60}, 5179 (1999).
A.~Lipowski and D.~Lipowska, Physica A {\bf 276}, 456 (2000).
\bibitem{Miller} A.~I.~Miller and M.~Foote, Science {\bf 302}, 1030 (2003).
\end{thebibliography}
\end {document}